\documentclass[twocolumn,showpacs,pra]{revtex4}
\usepackage{graphicx,subfigure,amsmath,amssymb,hyperref,mathrsfs,amsmath,amsfonts,amsthm,amssymb}
\usepackage{subfigure}

\usepackage{color} 
\usepackage{soul,ulem,cancel} 
\usepackage{epstopdf} 

\newcommand{\op}[1]{\hat{#1}}

\begin{document}

\title{{Extracting work from quantum states of radiation}}

\author{M. Kol\'a\v r}
\affiliation{ Department of Optics, Palack\'{y}
University, 771 46 Olomouc, Czech Republic}
\author{A. Ryabov}
\affiliation{
Charles University in Prague, Faculty of Mathematics and Physics, Department of Macromolecular Physics, V Hole{\v s}ovi{\v c}k{\' a}ch 2, 180~00~Praha~8, Czech Republic}
\author{R. Filip}
\affiliation{Department of Optics, Palack\'{y}
University, 771 46 Olomouc, Czech Republic}
\date{\today }

\begin{abstract}
Quantum optomechanics opens a possibility to mediate a physical connection of quantum optics and classical thermodynamics. We propose and theoretically analyze a one-way chain starting from various quantum states of radiation. In the chain, the radiation state is first ideally swapped to sufficiently large mechanical oscillator (membrane). Then the membrane mechanically pushes a classical almost mass-less piston, which is pressing a gas in a small container. As a result we observe strongly nonlinear and nonmonotonic transfer of the energy stored in classical and quantum uncertainty of radiation to mechanical work. The amount of work and even its sign depends strongly on the uncertainty of the radiation state.   
Our theoretical prediction would stimulate an experimental proposals for such optomechanical connection to thermodynamics. 

\end{abstract}
\maketitle

\section{Introduction}
\label{section-introduction}
Recent development of quantum optomechanics stimulates many proposals and experiments testing possibility of mechanical systems to mediate physical connections between physical platforms \cite{med1,med2,med3,med4,med5,med6,med7}. These physical connections bring together different physics and translate many physical notions beyond their mathematical analogies. Already from university studies, one would say that the most natural connection of classical mechanics is to classical thermodynamics \cite{greiner-textbook}. On the other hand, quantum optics is already directly connected to quantum mechanics by a pressure of light \cite{opmrev1,opmrev2} in many recent experimental achievements \cite{opm0,opm1,opm2,opm3,opm4,opm5,opm6,opm7}. It will be therefore natural to think about physical connection between quantum optics and classical thermodynamics mediated by mechanical systems. Theoretical thoughts about it and planning of future experiments are stimulated by recent fast progress in quantum optomechanics and many discussions about quantum thermodynamics in this context \cite{therm1,therm2,therm3,therm4,therm5,therm6,therm7,therm8,therm9}. 

Differently to many of these discussions, we focus on basic but important quantum-to-classical transition between quantum modes of light, mechanical oscillators at that border and fully classical system of the piston manipulating thermodynamic states of gas in the closed container \cite{greiner-textbook}. It is a kind of quantum-classical transition typical for the detectors registering light at quantum level. The mediating mechanical oscillator feels a position projection from the classical system and environment 
\cite{Zurek}. Quantum quadrature of light is therefore connected to mechanical position and further to the position of a classical piston. Altogether, it is a theoretical limit of one-way von Neumann chain, where one part drives the next one, but not vice versa. At this point, we simply ignore all the back actions, which can be the subject of further studies. Our first goal is to describe how inherent quantum uncertainty of the states of light translates into a classically uncertain position of the classical piston and classical average work on the ensemble of independently repeated experiments. It is a first gedanken experiment going towards more complicated and realistic considerations at this mechanical border between quantum and classical.

At the beginning of the challenging experiment, both classical and quantum uncertainty of virtual position and momentum quadrature variables of electric field of light or microwaves have to be translated to real uncertainty of the mechanical position and momentum of the mechanical system. Quantum optomechanics has already experimentally tested dedicated preparation and precise estimation of mechanical position with small uncertainty approaching quantum limits \cite{gr1,gr2,gr3,gr4,gr5,gr6,gr7,gr8,gr9}. Recently, quantum entanglement between radiation and mechanical systems and also quantum squeezing of mechanical uncertainty have been demonstrated \cite{opm3,opm5,opm6}. These experiments can be extended to prepare various quantum states of mechanical system \cite{st1,st2,st3,st4,st4a,st4b,st5,st6}. To reach a high quality of transfer for any state of light to mechanical systems, universal interfaces have been proposed as well \cite{int1,int2,int3,int4}. Remaining main experimental challenge is therefore the direct coupling of the mechanical system at quantum level to a classical thermodynamic system. A direct coupling between mechanical oscillators has been investigated in \cite{syn1,syn2,syn3,syn4}. Alternatively, mediated coupling between two mechanical systems, possibly representing the mechanical oscillator (or cantilever) and mechanical piston in future, has been proposed \cite{2mech0,2mech1,2mech2,2mech3,2mech4,2mech5,2mech6}.
Currently, progress in the research of levitating mechanical oscillators opens the door to such types of mechanical couplings \cite{lev1,lev2,lev3,lev4,lev5,lev6,lev7}.

All these points fully legitimate theoretical thoughts of a future experiment testing the simplest physical connection between quantum optics and classical thermodynamics. In this way, different forms of energy of light represented by thermal, coherent, squeezed and Fock states of light \cite{Scully} can be {\it physically} translated to thermodynamic work performed by a real device. It is a novelty for quantum optics, because classical and non-classical states of light can be seen from a perspective of classical thermodynamics through a specific physical chain. The simplest classical version of this chain (an interconnection between mechanics and thermodynamics) can be considered in terms of a mechanical cantilever whose single variable (mechanical position or momentum) drives a piston compressing an ideal gas in a solid container during a standard isothermal process. This process of energy transformation has interestingly nonlinear response, although the optomechanical coupling between light and mechanical oscillator remains typically linear. 

Considering an ensemble of such independent chains, classical or quantum uncertainty of light translates to an uncertainty of the piston position and finally to an uncertainty of work in classical isothermal process. In principle, this uncertainty can be present only on the ensemble, hence any individual chain in the ensemble does not principally fluctuate in time \cite{cerino}. The piston-gas-container can therefore follow the simplest isothermal process in standard thermodynamics \cite{greiner-textbook}. On the ensemble, the position of the piston and accessible work on testing body have to be however fully characterized by a statistical distribution over the ensemble. Classical average work on ensemble of the repeated experiments can be therefore seen as operational thermodynamic characterization of uncertainty present in the light. Beyond this basic level, analysis of the piston fluctuations driven by an environment can be added. In this sense, the piston can be understood as an over-damped particle under a Brownian motion \cite{cerino}. This can be viewed as an equivalent of the noise in a detector, which can affect the results of the detection process. It is an important test of stability of this chain under the fluctuations in time at the border of thermodynamics.

In this paper, we describe a gedanken but fully physical chain going from quantum optics to classical equilibrium thermodynamics. We investigate a position uncertainty of the piston and consequently, average work from isothermal process for several quadrature distributions of light. 
Our results show that the amount and even the sign of work obtained form the energy transfer through the chain depends non-trivially and nonmonotonically on the uncertainty of the radiation state. That is, contrary to what one can expect, the increase of uncertainty can cause either increase or decrease of the average output work.
In Sec.~\ref{section-description} we describe the components of our physical model of the chain. Sec.~\ref{section-equilibration} describes the approach of the piston subsystem to the thermal equilibrium. Sec.~\ref{section-work} describes the ensemble-averaged work as the partial outcome of the light to piston+gas energy transfer through the chain. Direct comparison for thermal, Fock, coherent, squeezed, and phase-randomized states of light is performed. 
The coherent states are the candidates for the most efficient average-energy-to-work transfer.


\section{Description of the Model}
\label{section-description}
\begin{figure}[htb]
\centering \hspace{-0.06\linewidth}
\includegraphics[width=.7\linewidth]{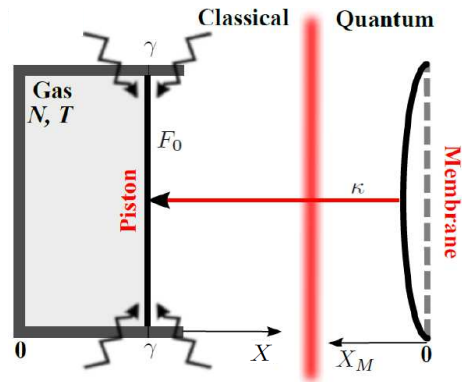}
\caption{Schematic of the physical model of the chain. Initially, the quantum state of the light is {\it ideally} transferred to the state of the mechanical membrane. This assumption causes the virtual absence of the light in our chain. The membrane position (together with its uncertainty) is transferred to the classical piston sealing the container with a classical ideal gas. 
} \label{figure-scheme}
\end{figure}

Initially, we assume that the quantum state of light mode is {\it ideally} transferred onto the state of the mechanical membrane. 
This fact causes virtual absence of the light mode in our model and allows for keeping our toy model consisting of three components only. Namely, the membrane subsystem, where the quantum projection to mechanical position happens, described quantum mechanically, the piston subsystem with stochastic description, and the classical non-fluctuating gas subsystem described in the thermodynamic language. 
Our model   therefore describes a one-way transfer of energy from the different quantum states of the membrane through the classically (and possibly stochastically) evolving piston, into the compression of the ideal gas stored in a container enclosed by the low-mass piston, Fig.~\ref{figure-scheme}. 
\subsection{The Membrane}
\label{subsection-membrane}

We describe the quantum part of the system, i.e., the membrane, by the quantum mechanical Hamiltonian of a harmonic oscillator
\begin{eqnarray}
\op{H}_{M}&=&\frac{\op{P}_M^2}{2m_M}+\frac{m_M\omega^2}{2}\op{X}_M^2,
\label{membrane-Ham}
\end{eqnarray}
where $X_M$ is measured from the membrane potential minimum. The quantum nature of $X_M$ can be omitted in the following, since we consider position distribution after perfect quantum projection of the membrane state by its environment, destroying any information about $P_M$. We use therefore only its position distribution. However, $X_M$ {\it does} represent the variable with a position uncertainty.
We assume the membrane to be coupled to a piston (described below) by the position-position type of coupling, typical for coupled harmonic oscillators. We consider this linear coupling to ensure that it cannot itself generate any nonlinear effects on the piston position. This type of coupling can be used for both quantum mechanical and classical oscillators. It reads
\begin{eqnarray}
{H}_{MP}=\kappa {X}_M {X}.
\label{membrane-piston-coupling}
\end{eqnarray}
This interaction Hamiltonian leads to the equation of motion for the piston position operator ${X}$ which contains the linear force $-\kappa {X}_M$ acting on the piston. 

The membrane is a conceptually important part of our gedanken experiment. It mediates the quantum-to-classical transition from the quantum state of light to the classical position distribution of the piston. The membrane undergoes a collapse in the position pointer basis on a time scale smaller than any other timescale assumed in the chain. After this projection the membrane description is effectively classical.
The direct transfer of quantum states of light onto the piston would be difficult to describe. 

\subsection{The Piston}
\label{subsection-piston}
As the next level of our chain we assume that the membrane acts on a very light microscopic piston. The one-way chain structure of our model implies that the back-action of the piston on the membrane can be neglected. 
In other words, we assume that the mass of the piston is much smaller compared to the membrane mass, $m_P\ll m_M$. This is the standard physical limit in which  the large membrane drives the smaller piston.
The piston represents a moving boundary used to compress a certain amount of a classical ideal gas in a cylinder. To test the basic robustness of the chain under fluctuations of the piston in time, we introduce stochastic Brownian motion of the piston caused by the piston environment. 
We assume that the dynamics of the piston position $X$ is described by the classical over-damped Langevin equation \cite{reif-textbook}
\begin{eqnarray}
\gamma \dot{X}=F(X,t)
+\sqrt{2\gamma k_BT}\,\xi (t),
\label{piston-Langevin}
\end{eqnarray}

Thus, we model the fluctuations of piston as if it would be a classical over-damped Brownian particle immersed in a heat bath of temperature $T$. 
In Eq.~(\ref{piston-Langevin}), $\gamma$ stands for the damping coefficient, $k_B$ is the Boltzmann constant, $F(X,t)$ is the non-fluctuating, but possibly uncertain, force acting on the piston, and $\xi(t)$ is the Gaussian white noise which accounts for the thermal fluctuations of the surroundings ($\langle\xi(t)\rangle =0$, $\langle\xi(t)\xi(t')\rangle=\delta(t-t')$). The force $F(X,t)$ is specified below in Eq.~(\ref{force}). The force uncertainty arises from the uncertainty of  mechanical system prepared by light. Such an approximate description of the dynamics is appropriate on the time scale $t \gg m_P/\gamma$, where $m_P$ is the mass of the piston.

\subsection{The Ideal Gas}
\label{subsection-gas} 
The third part of our model is the cylinder (container sealed by the piston) containing a certain amount ($N$ particles) of the classical ideal gas. The walls of the cylinder are kept at the constant temperature $T$.  Throughout the paper, we assume the particle number $N$ to be sufficiently low and  the heat bath temperature $T$ high. Under these assumptions the state of the gas is well described by the equation of state $PV=Nk_BT$, with $P$, $V=SX$ being the pressure and volume of the gas, respectively, with $S$ the piston cross-section and $X$ the piston position with respect to the bottom of the container. Thus, this equation can be readily recast into the form 
\begin{eqnarray}
PS=\frac{Nk_BT}{X}.
\label{gas-state-eq}
\end{eqnarray}
The left hand side of Eq.~\eqref{gas-state-eq} represents the pressure force the gas exerts on the piston, while the right-hand side, strongly nonlinear in piston position $X$, defines the piston position dependence of that force. Further we assume that the medium surrounding the gas container exerts a pressure force, $-F_0$, on the piston as well, allowing for establishment of mechanical equilibrium of the piston.

\section{Evolution of the Piston Towards Thermal Equilibrium}
\label{section-equilibration}
In this section, we describe the piston dynamics and characterize its equilibrium position distribution. To begin with, we specify the form of the force $F(X,t)$ in Eq.~\eqref{piston-Langevin}: 
\begin{eqnarray} 
F(X,t)=-\kappa X_M-F_0+\frac{Nk_BT}{X}.
\label{force}
\end{eqnarray}
The first right-hand side (RHS) term comes from the Hamiltonian~\eqref{membrane-piston-coupling}. 
We assume throughout the paper that $X_M$ changes at such time scale that it can be considered {\it constant} on the time scale of equilibration of the piston. Thus, we can consider an ensemble of equilibrating pistons whereas for each member of this ensemble the value of $X_M$ on the RHS of Eq.~\eqref{force} is constant and sampled from some probability density function (PDF). In other words, $X_M$ is {\it uncertain} (random) but not fluctuating variable, throughout the paper.  

The second and the third term in Eq.~\eqref{force} comes from Eq.~\eqref{gas-state-eq} and discussion below it. These terms describe the possibility the gas equilibrates mechanically with its surroundings. 
The Langevin equation for the piston position, Eq.~\eqref{piston-Langevin}, using Eq.~\eqref{force} as the RHS gets its final form
\begin{eqnarray}
\gamma \dot{X}=-\kappa{X}_M-F_0+\frac{Nk_BT}{X}
+\sqrt{2\gamma k_BT}\,\xi (t).
\label{piston-Langevin-final}
\end{eqnarray}
Equation~\eqref{piston-Langevin-final} is a stochastic differential equation describing the time dependence of the stochastic process $X(t)$. It takes simultaneously into account both the uncertainty of the membrane and the fluctuations of the piston. 
This equation is equivalent to the Fokker-Planck (F-P) equation for the PDF $\rho$ of the piston position $X(t)$, $\rho\equiv \rho(X,t)$. The F-P equation equivalent to Eq.~\eqref{piston-Langevin-final} has the form
\begin{eqnarray}
\nonumber
\frac{\partial \rho}{\partial t}=&-&\frac{1}{\gamma}\frac{\partial}{\partial X}\left[\left(-\kappa{X}_M-F_0+\frac{Nk_BT}{X} \right)\rho\right]\\
&+&\frac{k_BT}{\gamma}\frac{\partial^2 \rho}{\partial X^2}.
\label{piston-FP}
\end{eqnarray}

The deterministic forces on the RHS of Eq.~\eqref{piston-Langevin-final} acting on the piston can be derived from the potential
\begin{eqnarray}
V(X)=\left({\kappa{X}_M+F_0} \right)X-
{Nk_BT}\ln{(X/L)},
\label{piston-potential}
\end{eqnarray}
where $X_M$ represents the membrane position and $L$ is the integration constant making the argument of the $\ln(x)$ function dimensionless. We choose the value of $L$ such that it corresponds to the position of the potential minimum for $\kappa=0$, defining the length unit as 
\begin{eqnarray}
L\equiv Nk_BT/F_0.\label{length-unit}
\end{eqnarray}
With this definition, we can introduce dimensionless variable $x\equiv X/L$ and rewrite Eq.~\eqref{piston-potential} into the form
\begin{eqnarray}
V(x)=Nk_BT\left(\alpha x-
\ln{x}\right),\quad \alpha\equiv 1+\frac{\kappa X_M}{F_0}.
\label{piston-potential-dimensionless}
\end{eqnarray}

It is not possible to solve for time dependent $\rho(X,t)$ from Eq.~\eqref{piston-FP} analytically, but one can obtain the equilibrium position PDF of the piston $\rho(X)\equiv \lim_{t\rightarrow\infty}\rho(X,t)$ as the equilibrium Gibbs distribution 
\begin{eqnarray}
\nonumber
\rho(X)&=&\frac{1}{Z}\exp\left[-\frac{V(X)}{k_BT}\right],\\
Z&\equiv& \int_{-\infty}^{\infty}\exp\left[-\frac{V(X)}{k_BT}\right]{\rm d}X.
\label{piston-Gibbs}
\end{eqnarray}

Note, the equilibrium solution appears as the result of the diffusion process. If the diffusion can be neglected we obtain the case of the piston drifting to minimum of the potential under an external force with uncertainty of the mechanical membrane. We will go back to this simplification later. 

\subsection{The membrane without uncertainty}
\label{subsection-mechanics-deterministic}
In this subsection we  consider $X_M$ to be constant and derive the equilibrium PDF for the position of the piston in two cases. The first one comprises that the membrane is coupled to the piston, i.e., $\kappa\neq 0$, while the second one describes the situation with the two subsystems uncoupled, $\kappa =0$.

\begin{figure}[htb]
\centering \hspace{-0.06\linewidth}
\includegraphics[width=.9\linewidth]{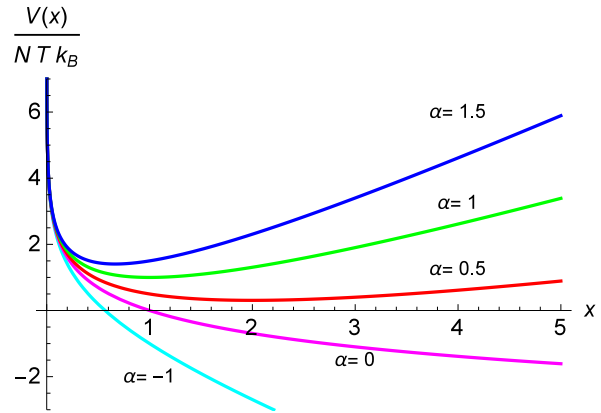}
\caption{An example of the behavior of the potential $V(x)/Nk_BT=\alpha x-\ln{x}$, for $\alpha=\{-1,0,0.5,1,1.5\}$. For $\alpha\leq 0$, no local minimum exists. For increasing $\alpha$, the potential is tighter and the position of its minimum shifts towards 0.
} \label{figure-potential}
\end{figure}

The nonlinear potential $V(x)$, Eq.~\eqref{piston-potential-dimensionless}, allows for the piston equilibration only if it possesses a local minimum with respect to $x$. This is equivalent to the constraint on the value of the membrane position, yielding $\alpha >0$. Taking the potential for $\alpha=1$ as a reference, we can deduce the following. For $X_M> 0$ ($\alpha> 1$), the membrane compresses the piston (the potential is  tighter). If $-F_0/\kappa <X_M<0$ ($0<\alpha<1$), the piston is expanded (the potential is more open). In the case $X_M\leq -F_0/\kappa$ ($\alpha\leq 0$) the potential has no local minimum, thus no equilibrium piston position distribution $\rho(x)$ exists, cf. Fig.~\ref{figure-potential}. 

Further, we always assume that the equilibrium PDF, Eq.~\eqref{piston-Gibbs}, for the piston position exists ($\alpha > 0$). In the dimensionless variables it is given by the Gamma distribution
\begin{eqnarray}
\label{piston-explicite-Gibbs}
\rho(x)&=&\frac{1}{Z} \, x^{N}\exp\left[ - \alpha Nx\right], \quad x\geq 0,\\
 Z&=& \frac{ \Gamma\left(N+1\right) }{
\left(\alpha N\right)^{\left(N+1\right)} }.
\end{eqnarray}
The mean $\overline{x}$ and the variance $\sigma^2=\overline{x^2}-\overline{x}^2$ of this distribution are given by
\begin{equation}
\overline{ x} = \frac{N+1}{\alpha N},\quad \sigma^{2} =\frac{N+1}{ (\alpha N)^2}.
\label{piston-explicite-if-Gibbs-deterministic}
\end{equation}

\begin{figure}[htb]
\centering \hspace{-0.06\linewidth}
\includegraphics[width=.9\linewidth]{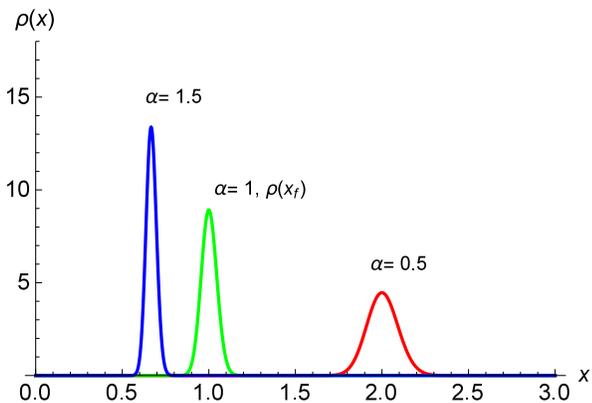}
\caption{Equilibrium position distribution, Eq.~\eqref{piston-explicite-Gibbs}, for $\alpha=\{0.5,1,1.5\}$ and $N=500$. For $\alpha =1$ the state is equivalent to the final $\rho(x_f)$ distribution. Note that due to the high value of $N$, the expected distribution skewness is not well distinguishable.
} \label{figure-piston-position-distributions}
\end{figure}

The description of the time sequence of one experimental run can be the following. Initially, the piston is coupled to the membrane, $\kappa\neq 0$, which is determining the value of $\alpha>0$, in the potential \eqref{piston-potential-dimensionless} during the whole equilibration process. After the waiting time long enough for the equilibration, the piston approaches the initial equilibrium position distribution, denoted $\rho(x_i)$, where the subscript ``$i$'' stands for ``initial.'' It is defined by the initial set of parameters $N$ and $\alpha$. 
From this initial distribution, we  perform a sufficiently slow (quasistatic) and isothermal transformation of the  piston state into the ``final'' state, labeled by the subscript ``$f$,'' of the  piston position distribution, denoted $\rho(x_f)$, by slowly decreasing $\kappa\rightarrow 0$, corresponding to $\alpha\rightarrow 1$. Thus, changing the value of $\kappa$ ($\alpha$) allows us to tighten/open the potential and therefore compress/expand the piston with respect to the reference state $\alpha =1$.
The position PDF, $\rho(x)$, is illustrated in Fig.~\ref{figure-piston-position-distributions} for different values of $\alpha$. 

Notice that for high values of $N$ higher statistical moments of $\rho(x)$ tends to be negligible, as evident from Fig.~\ref{figure-piston-position-distributions}. In the limit $N\gg 1$ the variances vanish as well 
 (cf. Eq.~\ref{piston-explicite-if-Gibbs-deterministic})
and the distributions approach
\begin{eqnarray}
\nonumber
\lim_{N\rightarrow \infty}\rho(x_i)&=&\delta(x_i-1/\alpha),\\
\lim_{N\rightarrow\infty}\rho(x_f)&=&\delta(x_f-1),\label{piston-thermodynamic-limit}
\end{eqnarray}
where $\delta(x)$ represents the delta function. The piston then always reaches the minimum of the potential without any fluctuations. The same result can be obtained, if the diffusion in Eq.~\eqref{piston-FP} can be neglected. 
These deterministic results are valid in the thermodynamic limit and describe the standard textbook situation of the piston with precise, non-fluctuating position.
The limit $N\to \infty$ is rather straightforward in the dimensionless coordinates. In physical units all three quantities $N$, $\kappa X_M$, $F_0$ become large and the latter two are proportional to $N$.

Using Eq.~\eqref{piston-explicite-if-Gibbs-deterministic}, we can determine, as well, the value of the mechanical Signal-to-Noise Ratio (SNR)  
of the state $\rho(x_i)$ as 
\begin{eqnarray}
{\rm SNR}=\frac{\overline{x}_i}{\sigma_i}=\sqrt{N+1},\label{equation-SNR}	
\end{eqnarray}
which is independent of $\alpha$ for fixed $N$. The interpretation of this result is such that, for fixed $N$, the change (increase/decrease) in the mean value $\overline{x}_i$ is always accompanied by the same change (increase/decrease) in the standard deviation $\sigma_i$. 
	
\subsection{The Membrane with Uncertainty}
\label{subsection-mechanics-fluctuating}
In this subsection we assume $X_M$ to take random values sampled from its PDF, corresponding to the fluctuations of $\alpha$ in Eqs.~\eqref{piston-potential-dimensionless}-\eqref{piston-explicite-if-Gibbs-deterministic}. These $\alpha$ fluctuations have to have the timescale separated with respect to the piston equilibration timescale, in the sense discussed bellow Eq.~\eqref{force}. Hence, the potential felt by the piston during each equilibration, Eq.~\eqref{piston-potential-dimensionless}, remains constant, as well as the form of the piston position distribution $\rho(x_i)$, Eq.~\eqref{piston-explicite-Gibbs}. 
Thus, the final form of the PDF $\rho(x_i)$ for the initial piston position will be the result of the averaging over the random values of $\alpha$, 
\begin{eqnarray}
\rho(x_i)&=&\frac{\int_{0}^\infty p(\alpha)Z_i^{-1}x_i^{N}\exp\left[ -\alpha Nx_i\right]{\rm d}\alpha}{\int_{0}^\infty p(\alpha){\rm d}\alpha}.
\label{piston-explicite-i-Gibbs}
\end{eqnarray}
The presence of the denominator in Eq.~\eqref{piston-explicite-i-Gibbs} ensures the correct normalization of the $\rho(x_i)$. This factor reflects the cutoff (neglection) of $p(\alpha)$ for $\alpha\leq 0$ because for these values no stationary piston state exists. It is because the piston position can be unstable and therefore, we have the probability
\begin{eqnarray}
p_s\equiv\int_{0}^\infty p(\alpha){\rm d}\alpha
\label{equation-success-prob}
\end{eqnarray}
of a successful experimental run less than unity. It is common for many conditional experiments with unstable systems. We can rectify the membrane position to be sure that equilibrium piston position will exist. All results from now on will be therefore conditioned by such rectifier. Note, the rectification and subsequently the success probability depends on $F_0$. It can be optimized to reach optimal regime of the transfer from mechanical uncertainty to thermodynamic average work.

Now, we can proceed to discuss the specific choice of $p(\alpha)$ in Eq.~\eqref{piston-explicite-i-Gibbs}. The most natural family of states of the membrane are the Gaussian states, characterized by the mean value $X_0$ and the variance $\overline{\epsilon}^2$. They characterize, e.g., the thermal as well as the coherently displaced squeezed ground states of the mechanical membrane.   
For the above mentioned Gaussian state we obtain from the original membrane position distribution $\rho(X_M)$ the induced $p(\alpha)$ distribution
\begin{eqnarray}
\nonumber
\rho(X_M)&=&\frac{1}{\sqrt{2\pi}\overline{\epsilon}}\exp\left[-\frac{\left( X_M-X_0 \right)^2}{2\overline{\epsilon}^2}\right]\Rightarrow \\
\nonumber
p(\alpha)&=&\frac{1}{\sqrt{2\pi}{\epsilon}}\exp\left[-\frac{(\alpha-\alpha_0)^2}{2{\epsilon}^2}\right],\\
&&{\epsilon}=\frac{\kappa \overline{\epsilon} }{F_0},\;\alpha_0=\left(1+\frac{\kappa X_0}{F_0} \right).
\label{fluctuating-force-distribution-gauss}
\end{eqnarray}

Below, we will recognize two cases for $p(\alpha)$. For the membrane coherent state ${\epsilon}$ is fixed to the standard position deviation of the mechanical ground state $\epsilon_0= \sqrt{\hbar/(2m\omega)}$ and $\alpha_0$ is a free parameter, representing the ``coherent'' shift of the mean value. In contrast, the thermal state is characterized by $\alpha_0=1$ is kept constant and ${\epsilon}$ is changed as a free parameter, reflecting the increase in fluctuations of the thermal state corresponding to $p(\alpha)$. 

\begin{figure}[htb]
\centering \hspace{-0.06\linewidth}
\includegraphics[width=.9\linewidth]{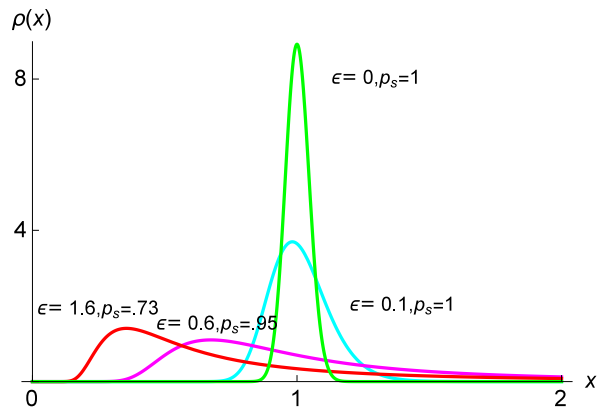}
\caption{The example of the probability distributions $\rho(x_i)$ and $\rho(x_f)$, $N=500$, for different values of ${\epsilon}$, cf. Eq.~\eqref{fluctuating-force-distribution-gauss} if Gaussian with $\alpha_0=1$ is assumed (thermal state). The success probabilities $p_s$, Eq.~\eqref{equation-success-prob}, are shown. We see that the increasing variance ${\epsilon}$ of $p(\alpha)$ distribution increases the variance of $\rho(x_i)$, positively skews the $\rho(x_i)$ distribution, and causes the shift of the mean value of $\rho(x_i)$ towards $0$.  As the integration in Eq.~\eqref{piston-explicite-i-Gibbs} suggests the resulting $\rho(x_i)$ is a weighted sum of Gamma distributions with different parameter $\alpha$. Distributions with small $\alpha$ have large mean value and variance creating fat tails of $\rho(x_i)$, whereas large $\alpha$ implies very narrow $p(\alpha)$ distributions with small mean values, causing the peaks of $\rho(x_i)$ moving to the origin.
} \label{figure-fluctuating-eps}
\end{figure}

The effect of changing ${\epsilon}>0$, keeping $\alpha_0=1$ (thermal state) can be seen on Fig.~\ref{figure-fluctuating-eps}. The distribution $\rho(x_i)$, Eq.~\eqref{piston-explicite-i-Gibbs}, is a weighted sum of the gamma distributions, Eq.~\eqref{piston-explicite-Gibbs}, with $p(\alpha)$ as the weights. The shapes of the resulting $\rho(x_i)$ on Fig.~\ref{figure-fluctuating-eps} result from their composition of the type of functions seen on Fig.~\ref{figure-piston-position-distributions}. The Gamma distributions with small $\alpha_i$ sum up to the thick distribution tales, while large $\alpha_i$ create the positively skewed maxima closer to $x=0$. 

\begin{figure}[htb]
\centering \hspace{-0.06\linewidth}
\includegraphics[width=.9\linewidth]{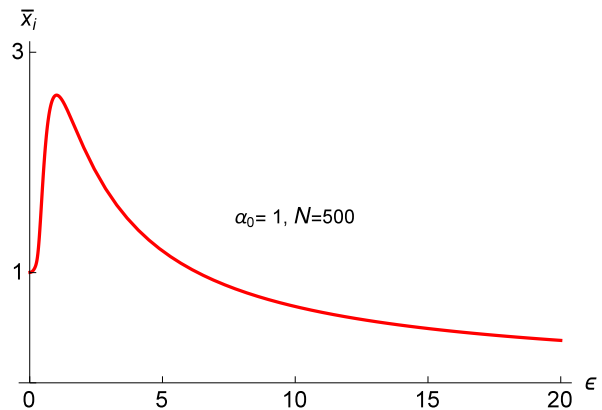}
\caption{The dependence of the mean value $\overline{x}_i$ from Eq.~\eqref{piston-explicite-i-Gibbs}, on the parameter ${\epsilon}$, Eq.~\eqref{fluctuating-force-distribution-gauss}, for fixed $\alpha_0=1$ and $N=500$. 
} \label{figure-fluctuating-eps-means}
\end{figure}

\begin{figure}[htb]
\centering \hspace{-0.06\linewidth}
\includegraphics[width=.9\linewidth]{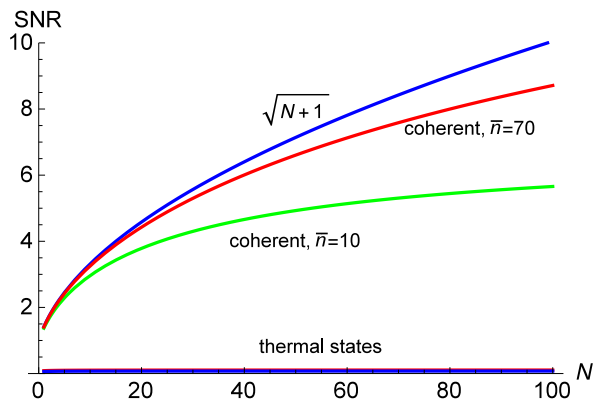}
\caption{The mechanical SNR plotted for different states \eqref{piston-explicite-i-Gibbs}, compared to the result \eqref{equation-SNR}. The $p(\alpha)$ distributions correspond to the coherent and thermal states, Eq.~\eqref{fluctuating-force-distribution-gauss}, with different values of the parameter $\overline{n}$, see Subsec.~\ref{subsection-performed-work}.
} \label{figure-fluctuating-eps-snr}
\end{figure}

One may naively tend to expect that while the modes (maxima positions) of the distributions $\rho(x_i)$, Eq.~\eqref{piston-explicite-i-Gibbs}, tend to decrease monotonically, Fig.~\ref{figure-fluctuating-eps}, so do their mean values. Figure~\ref{figure-fluctuating-eps-means} shows that this is not the case. Dependent on the parameter ${\epsilon}$, the mean value $\overline{x}_i$ increases initially above $\overline{x}_i=1$ (the mean value for the {\it final} distribution as well) and then falls back below this value. One may interpret such behavior as an increase / decrease of the average piston position with respect to the final position distribution and its $\overline{x}_f$. The SNR, Eq.~\eqref{equation-SNR}, for the piston states \eqref{piston-explicite-i-Gibbs}, using different $p(\alpha)$ distributions of the type \eqref{fluctuating-force-distribution-gauss}, are shown in Fig.~\ref{figure-fluctuating-eps-snr}. 

It shows that for intuitive understanding of the piston mechanics is better to keep in mind that the thermal energy of the membrane is transformed into the shift of the piston distribution maximum. Simultaneously, the local convexity around the maximum increases as well. It is therefore not easily predictable, how much average work can be obtained.  

\begin{figure}[htb]
\centering \hspace{-0.06\linewidth}
\includegraphics[width=.9\linewidth]{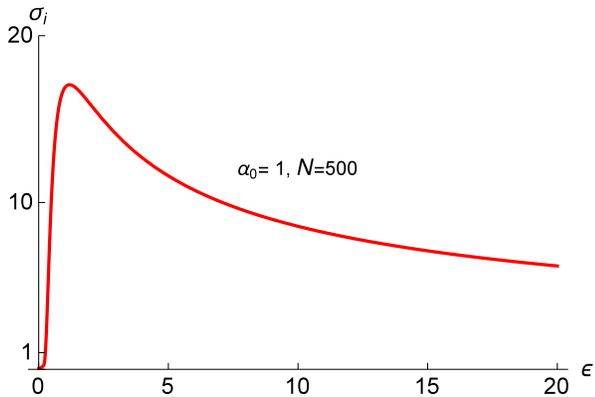}
\caption{The dependence of the standard deviation ${\sigma}_i$ from Eq.~\eqref{piston-explicite-i-Gibbs}, on the parameter ${\epsilon}$, Eq.~\eqref{fluctuating-force-distribution-gauss}, for fixed $\alpha_0=1$ and $N=500$. 
} \label{figure-fluctuating-eps-stdevs}
\end{figure}

\section{Work of the gas and piston}
\label{section-work}
This section examines the thermodynamic consequences of the results of nonlinear stochastic mechanics, derived in Sec.~\ref{section-equilibration}, first in the limit of large $N$ (with suppressed position fluctuations) and second with these fluctuations taken into account. 

In the previous section, we have understood the  mechanical motion of the piston with increasing uncertainty.  Our main goal in this section is therefore to quantify the average work done by the gas and piston expanding from the positions distributed with $\rho(x_i)$ into the positions distributed with $\rho(x_f)$ during the reversible isothermal expansion of the piston enclosing the ideal gas of $N$ particles and driven by the position of the membrane, $X_M$, distributed with $\rho(X_M)$, Eq.~\eqref{fluctuating-force-distribution-gauss}. We discuss examples how this average work depends on the type of the $\rho(X_M)$ distribution, hence, $p(\alpha)$ distribution, cf. Eq.~\eqref{fluctuating-force-distribution-gauss}. In this respect we consider an ensemble of reversible expansion/compression processes of an ideal gas in contact with the heat reservoir of the temperature $T$, Fig.~\ref{figure-scheme}, where the particular realizations of the ensemble are characterized by different values of the membrane position $X_M$, thus the parameter $\alpha$ determining the potential, Eq.~\eqref{piston-potential-dimensionless}. The ensemble average is taken over these $\alpha$ values. We point out that we {\it do not} take into account the work used to create the complete chain, i.e., the work necessary to map the state of the light onto the membrane and to establish the membrane-piston interaction. 
The reversibility condition can be satisfied by adiabatic switching off the piston coupling to the membrane, $\alpha\rightarrow 1$ in Eq.~\eqref{piston-explicite-Gibbs}. 

\subsection{The Thermodynamic Limit}
\label{subsection-thermodynamic-limit}
This subsection describes the results stemming from the limit of $N\gg 1$. This limit implies for the piston position distributions the form of Eq.~\eqref{piston-thermodynamic-limit}, represented by the delta functions. Further, we distinguish two regimes: (i) without any uncertainty in $X_M$, $\rho(X_M)$ being a delta function, (ii) with possible uncertainty in $X_M$.

First, let us analyze the deterministic case, i.e., the membrane being characterized by a sharp value of $X_M$, thus $\alpha$, without any uncertainty. 
In such case one can apply the results of macroscopic equilibrium thermodynamics describing the isothermal, reversible expansion (compression) of the ideal gas \cite{greiner-textbook} yielding the mechanical work, with the ideal gas pressure, Eq.~\eqref{gas-state-eq}
\begin{eqnarray} 
\nonumber
W&\equiv &-\int^{x_f}_{x_i}PS\;{\rm d}x=-Nk_BT\ln\frac{x_f}{x_i},\\
w(\alpha)&\equiv &\frac{W}{Nk_BT}=-\ln\alpha,\;\alpha=1+\frac{\kappa X_M}{F_0}. 
\label{thermodynamics-work}
\end{eqnarray}
Equation \eqref{thermodynamics-work} forms an important result of our thermodynamic analysis. It represents the (macroscopic) thermodynamic work done on the ideal gas, i.e., when $W$ is positive, the positive amount of work is done by an external agent on the ideal gas. In the thermodynamic limit ($N\gg 1$) the work $W$ has a sharp (deterministic) value for given $x_f$, $x_i$, $\alpha$. The normalized work is $w(\alpha)=-\ln\alpha$.

The second case described in this subsection has logic similar to the Subsec.~\ref{subsection-mechanics-fluctuating}. In this case, each thermalisation of the piston is realized in a different potential, Eq.~\eqref{piston-potential-dimensionless}, due to the uncertainty in the parameter $\alpha$, the slope of the thermalisation potential. 
Each member of the ensemble characterized by different value of $\alpha$, Eq.~\eqref{piston-potential-dimensionless}, is realized with the probability $p(\alpha)$, Eq.~\eqref{fluctuating-force-distribution-gauss}. For any $\alpha$-dependent physical quantity, we can observe its values only on this ensemble and represent such quantity by its mean value. This is the case for $w(\alpha)$, Eq.~\eqref{thermodynamics-work}, as well, leading to
\begin{eqnarray}
\overline{w}&=&-\overline{\ln\alpha}=-\int_{-\infty}^\infty\ln(\alpha)\overline{p}(\alpha){\rm d}\alpha,\;\label{thermodynamics-work-mean} \\ \nonumber
\overline{p}(\alpha)&\equiv &\frac{\theta (\alpha) p(\alpha)}{\int_{-\infty}^\infty \theta (\alpha)p(\alpha)\,{\rm d}\alpha},
\end{eqnarray}
where $\theta(\alpha)$ is the Heaviside step function ensuring the cut-off $\alpha>0$ necessary for the successful thermalization. 
The denominator of $\overline{p}(\alpha)$ represents the probability of successful experiment, Eq.~\eqref{equation-success-prob}. The average work $\bar{w}$ is therefore defined on an sub-ensemble of successful experimental runs. Equation~\eqref{thermodynamics-work-mean} yields the average work done by the piston during the isothermal reversible transition from $\lim_{N\rightarrow\infty}\rho(x_i)$, Eq.~\eqref{piston-explicite-i-Gibbs}, to $\rho(x_f)=\delta(x_f-1)$, Eq.~\eqref{piston-thermodynamic-limit}. For each value of $\alpha$ from the ensemble, $w(\alpha)$ yields a certain value of the mechanical work, but these values are sampled randomly from the distribution $p(\alpha)$. Equation~\eqref{thermodynamics-work-mean} is the functional relation $p(\alpha)\mapsto\overline{w}$, yielding $\overline{w}$ for given $p(\alpha)$. Note, if the diffusion process for the piston can be neglected we arrive to the same formula, Eq.~\eqref{thermodynamics-work-mean}, for the average work extractable from position distribution of the membrane in our physical chain.  

In order to evaluate $\overline{w}$, Eq.~\eqref{thermodynamics-work-mean}, we have to restrict ourselves to numerical results. As an example of the distribution $p(\alpha)$ we take again the Gaussian family of the membrane states, Eq.~\eqref{fluctuating-force-distribution-gauss}. 

\begin{figure}[htb]
\centering \hspace{-0.06\linewidth}
\includegraphics[width=.9\linewidth]{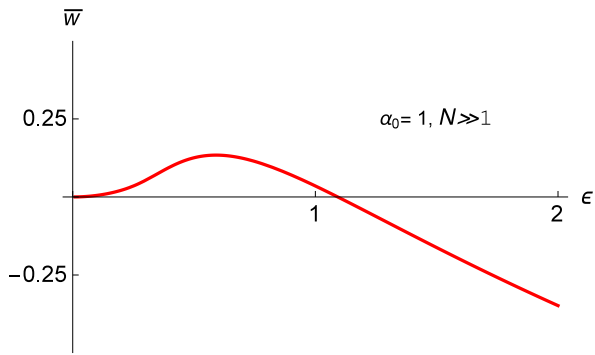}
\caption{Numerically obtained value $\overline{w}$ from Eq.~\eqref{thermodynamics-work-mean}. The independent parameter is ${\epsilon}=\kappa\overline{\epsilon}/F_0$, the standard deviation of the thermal state Eq.~\eqref{fluctuating-force-distribution-gauss}, its mean value $\alpha_0=1$ is fixed. 
} \label{figure-thermal-work-mean-value}
\end{figure}

\begin{figure}[htb]
\centering \hspace{-0.06\linewidth}
\includegraphics[width=.9\linewidth]{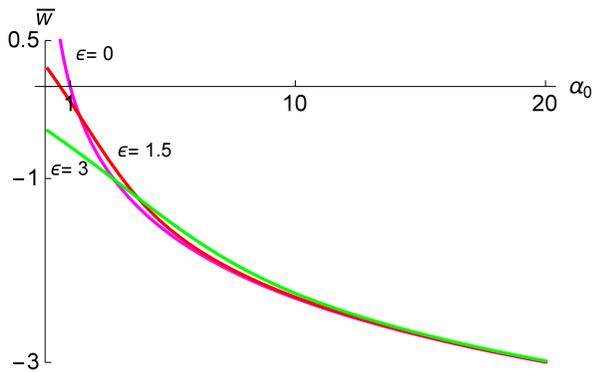}
\caption{Numerically obtained value $\overline{w}$ from Eq.~\eqref{thermodynamics-work-mean}. The independent parameter is the mean value $\alpha_0$ of the coherent state in Eq.~\eqref{fluctuating-force-distribution-gauss}. The different curves are parametrized by the values of variance $\epsilon=\overline{\epsilon}\kappa/F_0=\{0,1.5,3\}$ of this state. The behavior of $\overline{w}$ for $\alpha_0\approx 1$ is determined by $\epsilon$-dependence of $\overline{w}$ shown in Fig.~\ref{figure-thermal-work-mean-value}.  
Note the asymptotic behavior of all curves being the same, $\overline{w}\approx -\ln(\alpha_0)$. 
} \label{figure-coherent-work-mean-value}
\end{figure}

\subsection{The Piston as a Brownian Particle}
\label{subsection-thermodynamics-brownian}
This subsection describes the thermodynamic consequences for the work done in our scheme by the gas and fluctuating piston for the finite number of particles $N$. In this case, to correctly determine the average work, one has to naturally take into account the fluctuations of the piston position neglected in Subsec.~\ref{subsection-thermodynamic-limit}, which we do in a standard manner \cite{reif-textbook}. 

The piston equilibrium state $\rho(x;\alpha)$, Eq.~\eqref{piston-explicite-Gibbs}, is characterized by the mean potential energy $V(x;\alpha)$, Eq.~\eqref{piston-potential-dimensionless}, for the fixed parameter $\alpha$, 
\begin{eqnarray}
\overline{V}(\alpha)&=& \int_0^\infty V(x;\alpha)\rho(x;\alpha){\rm d}x,\\
\nonumber
\overline{v}(\alpha)&\equiv & \frac{\overline{V}(\alpha)}{Nk_BT},
\label{equation-thermo-average-energy}
\end{eqnarray}
in the overdamped regime assumed here. The infinitesimal dimensionless work, Eq.~\eqref{thermodynamics-work}, done by the piston is the infinitesimal change of average energy for the infinitesimal change of the parameter $\alpha$, 
\begin{eqnarray}
\delta w&\equiv &\frac{\partial\overline{v}(\alpha)}{\partial\alpha}{\rm d}\alpha\\
&=&\left.\int_0^\infty\frac{\partial v(x;\alpha)}{\partial\alpha}\right\vert_\rho\rho(x;\alpha){\rm d}x\;{\rm d}\alpha,
\label{equation-thermo-infi-work}
\end{eqnarray}
where the last partial derivative $\partial/\partial\alpha$ is taken under constant piston distribution condition. The total work done by the piston during the transition $\rho(x_i)\rightarrow\rho(x_f)$, see the discussion bellow Eq.~\eqref{piston-explicite-if-Gibbs-deterministic}, is 
\begin{eqnarray}
w(\alpha)=\int_\alpha^1 \delta w=-\frac{N+1}{N}\ln\alpha,
\label{equation-thermo-average-work}
\end{eqnarray}
yielding 
\begin{eqnarray}
\overline{w}=-\frac{N+1}{N}\overline{\ln\alpha}=-\frac{N+1}{N}\int_{-\infty}^\infty\ln(\alpha)\overline{p}(\alpha){\rm d}\alpha,\label{equation-thermo-average-work-micro}
\end{eqnarray}
resembling the results \eqref{thermodynamics-work},~\eqref{thermodynamics-work-mean} in the limit $N\gg 1$. The summand 1 in the nominator of \eqref{equation-thermo-average-work} reflects the fact that the piston itself, treated microscopically as a Brownian particle, performs average work on its surroundings. 

In the general case for $\overline{w}$, Eq.~\eqref{thermodynamics-work-mean}, i.e., the case when $\alpha$ is uncertain and distributed with $p(\alpha)$, the results in Figs.~\ref{figure-thermal-work-mean-value},~\ref{figure-coherent-work-mean-value} describe the behavior of $\overline{w}$ with high precision even in the case of non-negligible piston fluctuations, e.g., for $N\approx 10^3$. 

Equation~\eqref{thermodynamics-work-mean} also implies that for the states of the membrane with similar $p(\alpha)$, namely its statistical moments, the value of the mean work $\overline{w}$ ``extractable'' from $p(\alpha)$ is also similar. This is the case for, e.g., the thermal state of the membrane and the Fock state of the membrane, when both states posses the same  value of average energy. This fact is demonstrated in the next subsection for several possible states of the membrane.

\subsection{Work Performed by Different Membrane States}
\label{subsection-performed-work}
In this subsection we discuss the average work performance $\overline{w}$ of different mechanical membrane states. We examine the Gaussian family of states as an example, plus two representatives of non-Gaussian states. The reader may have the intuition about the results stemming from the Figs.~\ref{figure-thermal-work-mean-value},~\ref{figure-coherent-work-mean-value} already. The coherent and thermal states used here, are examples of the classical states of the mechanical membrane. Can the quantum mechanical coherence and non-classicality of the membrane state play any role in the work performance? As an example of the non-classical state we examine the harmonic oscillator eigenstate, the Fock state, the coherent squeezed state, and the phase-randomized coherent state \cite{Scully} as the distribution of the membrane positions. To make a fair comparison of the work performance, we parameterize all assumed states by the same  mean number of photons $\overline{n}$ they have. The mean photon number determines the average energy of the state, as well. 

The distribution of the {\it coherent state} is of the form \eqref{fluctuating-force-distribution-gauss}, with ${\epsilon}=1$ and $\alpha_0=1+2\sqrt{\overline{n}}$. The {\it thermal state} parameters are $\alpha_0=1$ and ${\epsilon}=\sqrt{1+2\overline{n}}$. 
\begin{figure}[htb]
\centering \hspace{-0.06\linewidth}
\includegraphics[width=.9\linewidth]{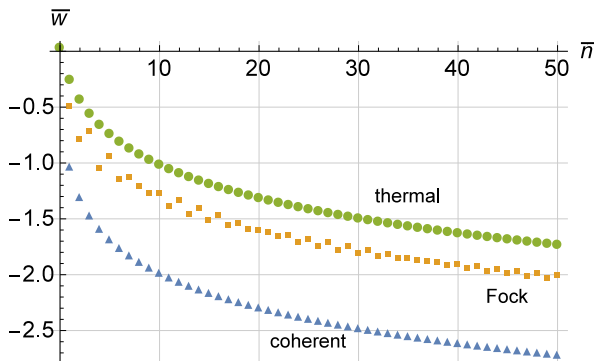}
\caption{Numerically obtained value $\overline{w}$ from Eq.~\eqref{thermodynamics-work-mean} for the subensemble of the successful thermalizations. The independent parameter is the photon mean number $\overline{n}$. The coherent states, number states, and thermal states with the same $\overline{n}$, i.e. the same average energy, are used at each point.
} \label{figure-coherentnumber-work-mean-value}
\end{figure}
\begin{figure}[htb]
\centering \hspace{-0.06\linewidth}
\includegraphics[width=.9\linewidth]{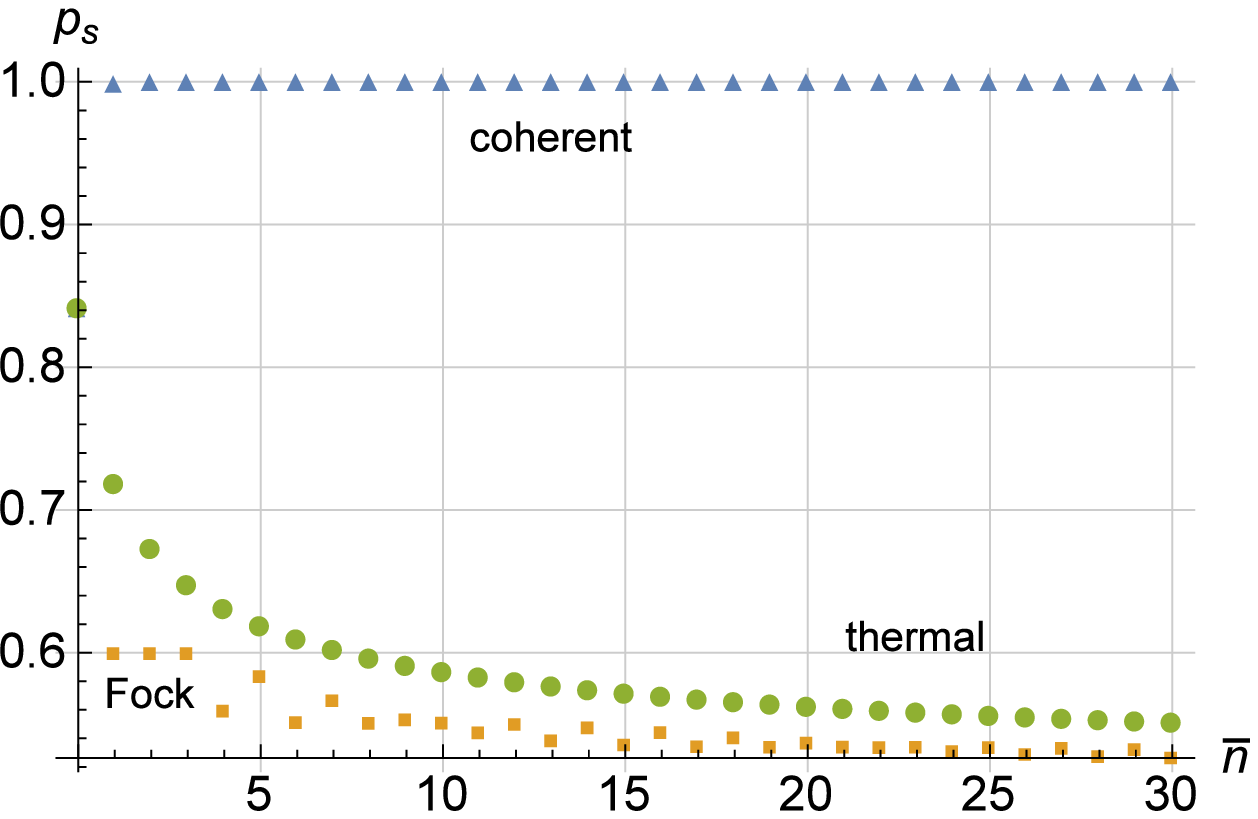}
\caption{Numerically obtained value of the experiment success probability $p_s$ from Eq.~\eqref{equation-success-prob}. The independent parameter is the photon mean number $\overline{n}$ for the states used in Fig.~\ref{figure-coherentnumber-work-mean-value}. 
} \label{figure-coherentnumber-psuccess}
\end{figure}
\begin{figure}[htb]
\centering \hspace{-0.06\linewidth}
\includegraphics[width=.9\linewidth]{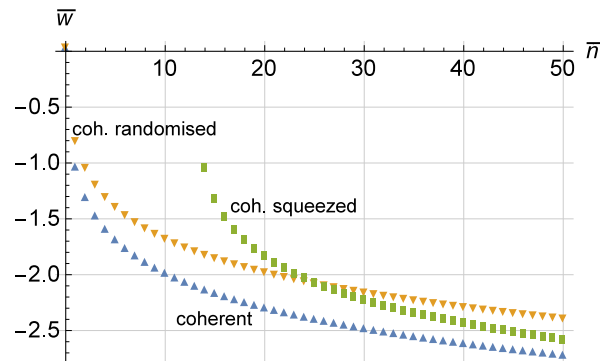}
\caption{Numerically obtained value $\overline{w}$ from Eq.~\eqref{thermodynamics-work-mean} for the subensemble of the successful thermalizations. The same variables as in Fig.~\eqref{figure-coherentnumber-work-mean-value} are shown for the same oscillator. The squeezing parameter $r=2$ for the squeezed state and the phase distribution variance $\tau=\pi/4$. 
} \label{figure-coherentsqueezed-work-mean-value}
\end{figure}
\begin{figure}[htb]
\centering \hspace{-0.06\linewidth}
\includegraphics[width=.9\linewidth]{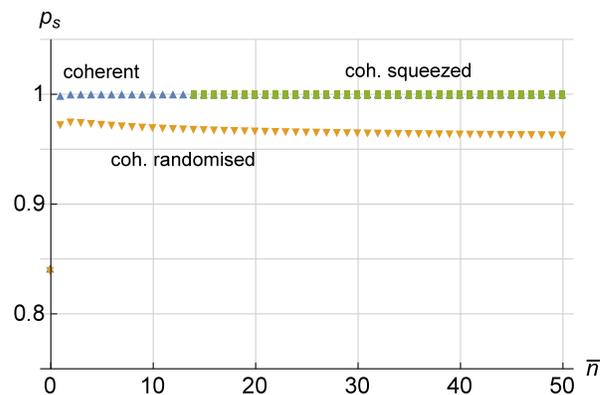}
\caption{Numerically obtained value of the experiment success probability $p_s$ from Eq.~\eqref{equation-success-prob}. The independent parameter is the photon mean number $\overline{n}$ for the states used in Fig.~\ref{figure-coherentsqueezed-work-mean-value}.
} \label{figure-coherentsqueezed-psuccess}
\end{figure}
Although the mean energy of each state mentioned is the same, the way how these states sample the value of $w(\alpha)=-\ln(\alpha)$, Eq.~\eqref{thermodynamics-work-mean}, through distribution $p(\alpha)$ is different. 

The coherent state can be typically envisaged like a peaked Gaussian distribution, $p(\alpha)$, with large mean value relative to its standard deviation. Thus, the coherent state samples the function $w(\alpha)=-\ln(\alpha)$ in a relatively narrow region of $\alpha$ values around $\alpha_0$, cf. Eq.~\eqref{fluctuating-force-distribution-gauss}. 
In comparison, the thermal state represents a Gaussian state with the vanishing mean value. Hence, the thermal state samples the function $w(\alpha)$ in a very wide region of $\alpha$. Because of the cutoff at $\alpha=0$, $p(\alpha)$ samples $w(\alpha)$ in a region $\alpha <1$, where the integral converges to a positive value, whereas in a region $\alpha >1$ the resulting integral is strictly negative and possibly divergent. These two contributions add, and for a wide enough distribution $p(\alpha)$ the negative part dominates. 

A similar situation appears in the case of the Fock state \cite{Scully}. It has formally the same standard deviation as the thermal state, even though it is an oscillating function with increasing amplitude for an increasing magnitude of the argument, in contrast to the thermal state. Due to this increase, the contribution to the negative part of the integral \eqref{thermodynamics-work-mean} increases faster in magnitude compared to the thermal state. This effect is resulting in the set of the Fock state data points located bellow the thermal state date points in Fig.~\ref{figure-coherentnumber-work-mean-value}.

Figure~\ref{figure-coherentsqueezed-work-mean-value} presents the results for other possible membrane states, namely the squeezed coherent state and phase-randomized coherent state \cite{Scully}.	 
The squeezed coherent state belongs to the family of the Gaussian states, Eq.~\eqref{fluctuating-force-distribution-gauss}, with  $\epsilon=\exp(-r)$ and $\alpha_0=1+2\sqrt{\overline{n}-\sinh^2(r)}$, $r$ being the squeezing parameter \cite{Scully}. The expression for $\alpha_0$ reflects the fact that squeezing is an active transformation changing the mean number $\overline{n}$ of photons with the squeezing parameter $r$. This has to be taken into account when comparing different state types with the same $\overline{n}$. In our example, squeezing reduces fluctuations of position distribution, one can therefore naively expect that they can produce more average work. We show the dependence of the average work $\overline{w}$ on the mean number $\overline{n}$ for the coherent-squeezed state with $r=2$.  

Last type of the state to be compared is the phase-randomized coherent state, belonging to the family of non-Gaussian states \cite{Scully}. In our example we use the phase-randomized coherent state with Gaussian phase distribution, namely
\begin{eqnarray}\nonumber
p(\alpha){\rm d}\alpha &=&\int_{-\infty}^\infty\frac{{\rm d}\phi}{2\pi\tau}\exp\left(-\frac{\phi^2}{2\tau^2}\right)\\
&\times &\exp\left[-\frac{\left(\alpha-(1+2\sqrt{\overline{n}}\cos\phi) \right)^2}{2}
 \right]{\rm d}\alpha,
\label{equation-phaserandomized}
\end{eqnarray}
where $\tau^2$ is the variance of the phase $\phi$ distribution. Phase randomization is broadening  the position distribution of coherent state, it is therefore important to check how sensitive is average work to phase instability of coherent state of the membrane. Figure~\ref{figure-coherentsqueezed-work-mean-value} shows the  phase-randomized coherent state with $\tau=\pi/4$. 

For both squeezed and phase-randomized states we note that the $\overline{w}$ values, Fig.~\ref{figure-coherentsqueezed-work-mean-value}, lie above the date for coherent state. In the asymptotic regime $\overline{n}\gg 1$ the approximate value of the integral \eqref{thermodynamics-work-mean} for the coherent states is
\begin{eqnarray}
\nonumber
\overline{w}_{\rm coh}\approx -\ln \alpha_0=-\ln 2-\frac{1}{2}\ln\overline{n},\;\overline{n}\gg 1,
\label{performance-mean-work}
\end{eqnarray}
placing the coherent state into the candidate-position for the state with the most effective energy transfer through the assumed chain. 
If one is interested in the work done on the {\it whole} ensemble, it is necessary to multiply the subensemble work values, Figs.~\ref{figure-coherentnumber-work-mean-value},~\ref{figure-coherentsqueezed-work-mean-value}, by the respective success probabilities $p_s$, Eq.~\eqref{equation-success-prob}, plotted on Figs.~\ref{figure-coherentnumber-psuccess},~\ref{figure-coherentsqueezed-psuccess}. This favors the coherent states again. 

Thus, within the limits of our scheme, the non-classicality of the states does not seem to represent any considerable advantage with respect to the work performance. It is a very interesting and unexpected result, considering complexity of the nonlinear way to obtain the average work from the position distribution.

\section{conclusions}
\label{section-conclusions}
We have presented a physical model of a von Neumann one-way chain consisting of the following parts. The (i) quantum state of radiation swapped to (ii) the quantum mechanical membrane (oscillator), linearly coupled to a (iii) one-dimensional piston sealing a certain number of (iv) classical ideal gas particles. The gas can be compressed or expanded as an effect of the membrane pressure exerted through the mutual coupling on the piston. This process is assumed to take place at constant temperature, due to the presence of a heat bath.

The mechanical effect of the membrane on the piston position distribution under the isothermal conditions is studied for different  position distributions of the membrane. The result is comprehensively expressed by the formula \eqref{equation-thermo-average-work-micro} for rectified distribution of membrane position ensuring existence of the equilibrium state. It can be simply used to operationally determine achievable thermodynamic work corresponding to any position distribution of the membrane or any other mechanical object.  

For an ensemble of the membrane positions, we determine the average work done by the gas and piston when isothermal reversible process of switching off the membrane-piston coupling is assumed. The average reversible work done by the gas isothermal transformation from the initial into the final states is compared for different membrane position distributions. Namely, the work is compared for some states from the Gaussian family of the membrane states, for its Fock states, and for the non-Gaussian phase-randomized coherent state. 

From our results, we conclude several facts. 
We observe non-trivial, strongly nonlinear and nonmonotonic, relationship between quantum and classical uncertainty of light and average work.  We observe clear advantage of Gaussian coherent states of light over thermal states, squeezed states or even highly non-classical Fock states of light to transform their energy into average work. We confirm that for large energy of light we asymptotically reach classical limit. The coherent states are the candidates for the most efficient  mean energy to work transfer among the states assumed in our work. As a goal of our future work we plan to examine the effect of finite time (irreversible) dynamics of the piston and hence the work distribution and irreversible-work losses. 

All these theoretical  results are important for further, more realistic, development of physical interconnection between quantum optics, optomechanics and classical thermodynamics.

\section*{Acknowledgments}
M.K. and R.F. acknowledge the support of the project GB14-36681G of the Czech Science Foundation.

\end{document}